\documentclass[a4paper,prd,amsfonts, amssymb, amsmath, reprint, showkeys, nofootinbib, twoside, superscriptaddress,onecolumn,notitlepage]{revtex4-1}
\usepackage[english]{babel}
\usepackage[utf8]{inputenc}
\usepackage[colorinlistoftodos, color=green!40, prependcaption]{todonotes}
\usepackage[pdftex, pdftitle={Article}, pdfauthor={Author}]{hyperref}
\usepackage{xcolor}
\usepackage{graphicx}
\usepackage{adjustbox, float}
\usepackage{placeins}
\usepackage[T1]{fontenc}
\usepackage{lipsum}
\usepackage{csquotes}
\usepackage{amsmath, amssymb, slashed, minitoc, amsthm,graphicx,caption, tensor, mathtools,wrapfig,bm,esint,subcaption,cancel,booktabs}
\usepackage{stackengine}

\usepackage{CJK}
\begin{document}
\title{Non-minimal Scalar Effective Cosmology}
\author{Callum L.\ Hunter}
\email{c.l.hunter@soton.ac.uk}
\affiliation{
Mathematical Sciences and STAG Research Centre, University of Southampton,\\
Highfield, Southampton, SO17 1BJ, UK}
\author{Michael Kenna-Allison}
\email{m.kenna-allison@soton.ac.uk}
\affiliation{
Mathematical Sciences and STAG Research Centre, University of Southampton,\\
Highfield, Southampton, SO17 1BJ, UK}

\date{\today} 
\begin{abstract}
\noindent
In this work we investigate the most general non-minimally coupled $\mathbb{Z}_2$ symmetric scalar-tensor effective field theory (EFT) of gravity up to dimension six in the operator expansion. The most general action is presented along with its equations of motion both in the covariant form and also in the coordinate form resulting from an FLRW analysis. The pressure and density of the scalar field are found as well as the equation of state parameter and some cosmological parameters. We analyse the background evolution of the scalar field within the framework of the slow-roll approximation and provide a brief discussion of $\mathbb{Z}_2$ symmetry breaking of the scalar field which gives an order of magnitude constraint on one of the coupling constants. To provide a concrete example we choose a couple of potentials for the scalar field and explore the cosmology. Some brief comments on the range of validity of the EFT are also offered as well as the connections between the general model and well-known examples in the literature. 
\end{abstract}
\maketitle
\section{Introduction} \label{intro}
\noindent
General relativity (GR) has been a successful theory for over 100 years and continues to accurately predict phenomena on both planetary and stellar scales - however there are issues with the theory, and these issues must be dealt with if we wish to understand the true nature of gravity. There a number of theoretical reasons why GR is an inadequate theory and the main issue is that it is non-renormalizable at more than one-loop \cite{oneloop1,oneloop2} and the issue becomes more acute when coupling to matter \cite{matter1,matter2}. GR and normal matter also fail to explain current cosmological data (including late-time inflation) and are unable to properly account for galactic rotation curves.

In order to address the late-time inflation of the universe, as suggested by Supernova data \cite{Supernova1, Supernova2, Supernova3,Supernova4}, there has been a flurry of activity concerning theoretical investigations of various models. A great number of models have been proposed including a dynamical cosmological constant \cite{DynamicLambda}, cosmic strings \cite{Cosmicstrings} and more recently giving the graviton a mass via the well-known dRGT model \cite{dRGT1,dRGT2} which has a generalised extension allowing for a stable cosmology \cite{Stablecosmo}. There have also been a number of studies using baryonic acoustic oscillations in the cosmic microwave background that have confirmed that baryonic matter makes up a small percentage of the total matter content of the universe \cite{BAO1,BAO2}. These facts point to some form of dark energy field; typically this field is described by modifying the minimal contributions to the matter part of GR, often via a new scalar field \cite{Minimal1,Minimal2} or by geometrically altering Einstein's equations leading to so-called modified gravity theories (for some reviews see \cite{mod1,mod2,mod3,mod4}). A subset of such theories are so-called non-minimally coupled scalar-tensor theories (STTs) in which the additional fields are directly multiplied into the curvature parameters appearing in the Einstein-Hilbert action \cite{scalartensor1,scalartensor2,scalartensor3,scalartensor4,scalartensor5}. It is quite natural to consider such theories since these kinds of couplings occur as a result of string theory which give rise to dilatonic couplings \cite{dilaton}. The first kind of non-minimally coupled theory to be studied for its cosmological applications was the well-known Brans-Dicke (BD) theory \cite{BD1,BD2,BD3,BD4} which recovers GR in the large limit of its novel coupling. There are numerous non-minimal STTs and for a comprehensive review of such theories see \cite{fujii1,STT1}.

In this paper we are interested in a similar approach which relies on the idea of a bottom-up effective field theory (EFT) \cite{Donoghue1,Burgess1,speedofgravity} with a non-minimally coupled scalar field. The premise of such an approach is to begin with a field theory that we know is experimentally sound - the Einstein Hilbert action. Then we supplement this action order by order in the mass dimension by adding higher order curvature terms. These terms must then be suppressed by some massive scale in order to make sure that the operators have the correct dimension, luckily GR already has such a mass scale\footnote{In fact it is this mass scale that leads to the issues of non-renomalizability.} in the form of the Planck mass squared $M_{Pl}^2$. There are a number of operators that one can add on and in fact the first order action has already been investigated in \cite{scalartensor1} which contained a $\phi^2R$ term. However, owing to the presence of a mass term, all of these higher order operators are non-renormalizable and hence the theory we describe in this paper is only valid up to some energy scale which is unknown but of order of the Planck mass. Hence the theory investigated in the present paper is simultaneously conservative and general in its implications. It is conservative since we will essentially consider only small deviations from the usual GR cosmology (although the results in this paper can be extended to general STTs by treating the results as exact rather than a perturbative expansion in the cut-off mass); but it is general as well shall allow all terms, including non-minimally coupled ones, such that the action obeys a scalar field $\mathbb{Z}_2$ symmetry. This symmetry is natural as it allows the scalar field to have stable vacua, it also allows for a vacuum expectation value which will induce a cosmological constant if one desires. The premise of this work is \`a la Weinberg \cite{Weinberg}, however in this work we ensure the scalar field has mass-dimension one and hence we take a careful accounting of mass dimensions in the effective expansion of the field theory. This point is subtle but important as it allows us to write down very explicitly the most general polynomial expansions of the scalar functions in \cite{Weinberg} in terms of the cut-off mass.

This article is laid out in four further sections. In Section \ref{Model} we lay out, what we believe to be, the most general non-minimal effective action up to mass dimension six in the curvature operators corresponding to order $M^{-4}$ in the cut-off mass of the EFT. Following this in Section \ref{Equations} we find the background field equations for the FLRW metric as well as the pressure and density of the scalar field. We also discuss some cosmological parameters. In Section \ref{Inflation} we investigate two common regimes in the literature namely the slow-roll model and the \textit{age of the universe model}. Finally, in Section \ref{conclusions} we draw our conclusions and lay out the structure of upcoming work as well as further ideas for the future.

\section{The Model}\label{Model}
\noindent
In this section we layout the form of the effective action up to dimension-6 in the operators; this action is the most general non-minimally coupled action up to equations of motion - the equations of motion will generate further terms that we do not consider here but they are essentially field redefintions and so we need not explicitly consider them here. We do not include a cosmological constant to the action since the scalar field potential automatically includes such a term if we allow for spontaneous symmetry breaking of the $\mathbb{Z}_2$ symmetry. Breaking this symmetry is subtle as the vacuum expectation value can be both constant and dynamical since the non-minimally coupled terms introduce curvature tensors to the potential of the scalar field. As a result, there is the possibility of back reaction from the background which affects the vacuum expectation value of the scalar field. We will not explore this explicitly here but it is discussed in \cite{Gorbar:2003yp}, as such we will assume for the most part that the vacuum expectation value is constant - although we do briefly discuss the restrictions this assumption places on the theory in later subsections.

Using the aforementioned restrictions the most general action we investigate, inspired in part by \cite{Weinberg}, is given by,
\begin{widetext}
\begin{equation}\label{CA}
    \begin{aligned}
    S=M^2\int d^4x\sqrt{-g}\Bigg(&R+\frac{1}{M^2}[\partial_\mu\phi\partial^\mu\phi+\xi_1\phi^2R]+\frac{1}{M^4}[\beta_1\phi^3\Box\phi+\beta_2(\Box\phi)^2+\partial^\mu\phi\partial^\nu\phi (\xi_2R_{\mu\nu}+\xi_3g_{\mu\nu}R)\\
    &+\phi^2(\xi_4R_{\mu\nu}R^{\mu\nu}+\xi_5R_{\mu\nu\rho\sigma}R^{\mu\nu\rho\sigma}+\xi_6 R^2)]-V(\phi)+\frac{\Omega^{(4)}}{M^2}+\frac{\Omega^{(6)}}{M^4}\bigg),
    \end{aligned}
\end{equation}
\end{widetext}
where $V(\phi)$ would typically represent the polynomial expansion\footnote{We say typically because in an effective expansion this is how we would denote the scalar potential, for example, to this order $V(\phi)\sim\phi^2+\phi^4+\phi^6$, with correct factors of $M$ inserted. However, we do not include this explicit form of the potential as it allows the results of later sections to be general - such as allowing for exponential potentials.} of $\phi$ in terms of the cut-off parameter $M$ and $\Omega$ are the higher-order curvature-only terms which can be found in Appendix \ref{appA} and \cite{speedofgravity,Dim6ops}. The $\beta_1$ and $\beta_2$ terms appear as though they can be ignored by using the equations of motion, however there are missing non-minimal terms which are generated once the equations of motions for $\phi$ are substituted into these terms. Hence this action represents the most general action up to equations of motion and field redefinitions - note that throughout this paper we work in the Jordan frame and do not use the Einstein frame. We choose the scalar field-only terms over the non-minimal terms as they make later analysis much easier. They also make the Einstein Fields equations (EFEs) simpler. The equations of motion are easily found, but are quite tedious. To find the field equations we employ the \textit{xTras} \cite{xTras} package from \textit{xAct} in \textit{Mathematica}. The scalar field equation of motion is given by,
\begin{widetext}
\begin{equation}
    \begin{aligned}\label{CSFE}
    \Box\phi+\frac{1}{2}\frac{\partial V}{\partial \phi}=&\xi_1\phi R+\frac{1}{M^2}\Bigg[\phi(\xi_4R_{\mu\nu}R^{\mu\nu}+\xi_5R_{\mu\nu\rho\sigma}R^{\mu\nu\rho\sigma}+\xi_6 R^2)+\Box\phi(3\beta_1\phi^2-\xi_3R)\\
    &+\nabla^\mu\phi(3\beta_1\phi\nabla_\mu\phi-\xi_3\nabla_\mu R-\xi_2\nabla^\nu R_{\mu\nu})+\beta_2\Box^2\phi-\xi_2\nabla^\mu\nabla^\nu\phi R_{\mu\nu}\Bigg],
    \end{aligned}
\end{equation}
\end{widetext}
note that this equation only has terms up to $M^{-2}$ since the scalar field has mass dimension 1 and all scalar terms are thus suppressed by $M^2$.
The modified EFEs are easily computed in \textit{Mathematica} - albeit they are rather complicated and opaque - and are given by,
\begin{equation}\label{CEFE}
    M^2G_{\mu\nu}=\Phi_{\mu\nu}+\Xi_{\mu\nu}+\Omega_{\mu\nu},
\end{equation}
where $G_{\mu\nu}$ is the Einstein tensor. We split the field equations into three parts for ease of display and also to separate out each of the types of terms that contribute to the action. Firstly, $\Phi_{\mu\nu}$ is the modified scalar field stress-energy tensor for the minimally coupled terms only and is given by,
\begin{widetext}
\begin{equation}\label{miniSET}
    \begin{aligned}
    \Phi_{\mu\nu}=\frac{1}{2}g_{\mu\nu}\bigg(&-V(\phi)+\partial_\rho\phi\partial^\rho\phi-3\frac{\beta_1}{M^2}\phi^2\partial_\rho\phi\partial^\rho\phi-\frac{\beta_2}{M^2}(\Box\phi)^2-\frac{2\beta_2}{M^2}(\partial_\rho\phi\Box\partial^\rho\phi-\partial_\rho\phi\partial_\sigma\phi R^{\rho\sigma})\bigg)\\
    &-\partial_\mu\phi\partial_\nu\phi+\frac{3\beta_1}{M^2}\phi^2\partial_\mu\phi\partial_\nu\phi+\frac{\beta_2}{M^2}(\partial_{(\mu|}\phi\Box\partial_{|\nu)}\phi-R_{(\mu|\sigma}\partial_{|\nu)}\phi\partial^\sigma\phi),
    \end{aligned}
\end{equation}
\end{widetext}
where $(|...|)$ indicates symetrization of the indices without weight, for example $(\mu|\nu)=\mu\nu+\nu\mu$. Note that the extra factor of -3 for the $\beta_1$ terms is due to integration by parts. The main thing to note in \eqref{CSFE} and \eqref{miniSET} is that these equations are highly non-trivial and non-linear, this is of course no surprise given the nature of non-minimally coupled theories and it is one of the main obstacles to performing analysis.

The non-minimal stress-energy tensor for scalar field terms is given by $\Xi_{\mu\nu}$ and $\Omega_{\mu\nu}$ is the stress-energy tensor from the higher order curvature operators. We choose not to display these two terms as their form is particularly unpleasant and the covariant equations do not add anything useful to the rest of the discussion. Such terms can easily be found by using \cite{xTras} in \textit{Mathematica} - there is an attached notebook which will generate such terms.

\
\section{The Background Equations}\label{Equations}
\noindent
In this section, and what follows, we shall assume that we are working in a spacetime where $k=0$, that is a flat spacetime. This is mainly motivated by the fact that the observable universe appears to have $k\sim0$. To find the equations of motion we work in the mini-superspace formalism. This is done by assuming that the metric may be described by a FLRW line element given in spherical coordinates by,
\begin{equation}\label{FLRW}
    \text{d}s^2=-n^2\text{d}t^2+ a^2\text{d}r^2+a^2r^2\text{d}\theta^2+a^2r^2\sin^2(\theta)\text{d}\varphi^2,
\end{equation}
where $a$ is the scale factor and $n=\text{d}\tau/\text{d}t$ is the lapse function for $\tau$ proper time and $t$ physical time. This metric slices the spacetime into spatial slices with zero curvature and this splitting will be exploited to find the equations of motion for $H$ and $\phi$ in the background in an efficient manner. 

One can now substitute \eqref{FLRW} into \eqref{CA} and assume a spatially homogeneous time-varying scalar field $\phi$ - this physically interprets the scalar field as making the couplings of the non-minimal terms dynamical and thus allows the `strength' of the gravitational force to be dynamic. Thence one obtains the mini-superspace action \cite{Stablecosmo}, and in doing so one finds the action,
 \begin{equation}\label{MSSA}
     \begin{aligned}
     S_{\text{MSS}}= \text{M}_{\text{PL}}^2 &\int  dt\; n\; a^3 \Bigg[  \frac{6}{n^2}\left(2H^2 + \dot{H}-\frac{n^2}{6}V(\phi)-\frac{ H \dot{n}}{n}\right) +\frac{6}{n^2 M^2}(\mathcal{S}_1+\mathcal{S}_2) + \frac{1}{ n^2 M^4}(\mathcal{S}_3 + \mathcal{S}_4 + \mathcal{S}_5)\Bigg]\\
     \end{aligned}
 \end{equation}
where $H=\dot{a}/a$ and we define the higher order parts of the action in the following manner,
\begin{equation}
    \begin{aligned}
    \mathcal{S}_1 =& 6 \tilde{\alpha}_2 \frac{H^4}{n^2} + 2 \tilde{\alpha}_2 \frac{\dot{H}^2}{n^2}- 6 \tilde{\alpha}_2 H^3 \frac{\dot{n}}{n^3} -4\frac{H}{n^3} \tilde{\alpha}_1 \dot{H}\dot{n} + 2 \frac{H^2}{n^4}\left(3 \tilde{\alpha}_2 n^2 \dot{H} + \tilde{\alpha}_1 \dot{n}^2 \right),\\
    \mathcal{S}_2 =& \xi_1 \bigg(\phi^2 \dot{H} + 2\xi_1\phi^2H^2  -\frac{ \phi^2H\dot{n}}{n}\bigg)-\frac{\dot{\phi}^2}{6},\\
    \mathcal{S}_3 =& \frac{1}{n^2}\Big[12 \left(\tilde{\xi}_4 H^4 + \tilde{\xi}_3\dot{H}^2 \right)\phi^2  - 3 \tilde{\xi}_2 \dot{H}\dot{\phi}^2 + 3 H^2 \left(4 \tilde{\xi}_4 \phi^2 \dot{H} - \tilde{\xi}_1\dot{\phi}^2  \right)\Big] +\\
    &\frac{\dot{n}}{n^3}\Big(12 \tilde{\xi}_4 H^3 + 3 H \tilde{\xi}_3 \dot{H}) \phi^2 - \tilde{\xi}_2 \dot{\phi}^2 \Big)  + \frac{12 \tilde{\xi}_4 H^2 \phi^2\dot{n}^2}{n^4},\\
    \mathcal{S}_4 =&  -3 \beta_1 \left(3 H \phi^3 \dot{\phi} + \phi^3 \ddot{\phi} \right) + \frac{\beta_1 \dot{n} \phi^3 \dot{\phi} }{n} +9\frac{\beta_2}{n^2}\dot{\phi}^2 + 3\beta_2H^2 ( 6H\dot{\phi} + \ddot{\phi}) \ddot{\phi}\\
    &+2\frac{\beta_2\dot{\phi}^2\dot{n}}{n^3}+ 2 \beta_2 \dot{\phi}\ddot{\phi} +\frac{\beta_2\dot{\phi}^2\dot{n}^2}{n^4},\\
    \mathcal{S}_5 =& \Omega_{MSS}^{(6)}.
    \end{aligned}
\end{equation}
We define the $\tilde{\xi}_i$ by,
 \begin{equation}
     \begin{aligned}
     \tilde{\xi}_1&=\xi_2+4\xi_3\\
     \tilde{\xi}_2&=\xi_2+2\xi_3\\
     \tilde{\xi}_3&=\xi_4+\xi_5+3\xi_6\\
     \tilde{\xi}_4&=3\xi_4+2(\xi_5+6\xi_6),
     \end{aligned}
 \end{equation}
and the $\tilde{\alpha}_i$ are given by,
\begin{equation}
    \begin{aligned}
    \tilde{\alpha}_1&=3\alpha_1+\alpha_2\\
    \tilde{\alpha}_2&=4\alpha_1+\alpha_2.
    \end{aligned}
\end{equation}
The dimension-6 contribution is given in Appendix \ref{appA} - and can be expressed in a relatively compact form. We can then vary the action in \eqref{MSSA} with respect to $\phi$, $n$ and $a$ in order to obtain the component equations of motion and thus deduce the pressure and density of the scalar field. In order to do this we adopt cosmological time and so throughout this section, which equates to setting $n=1$ after the relevant variations.

The $a$ equation of motion gives,
\begin{equation}\label{aeom}
    \begin{aligned}
    &\Theta = - \frac{V(\phi)}{2M^2} - \frac{P_1}{M^2} - \frac{P_2+P_3}{M^4} - (\Omega^{(6)}_{MSS})_{,a}
    \end{aligned}
\end{equation}
where,
\begin{equation}
    \Theta=2\dot{H}+3H^2,
\end{equation}
and
\begin{equation}\label{pressures}
    \begin{aligned}
    P_1 =& \xi_1 \left( \Theta \phi^2 + 2 \dot{\phi}^2 + (4 H \dot{\phi} + 2 \ddot{\phi})\phi   \right) + \tilde{\alpha}_1 \left( 18 \dot{H}^2 + 36 H^2 \dot{H} + 24 H \ddot{H} + 4 \ddot{H} \right) - \frac{1}{2}\dot{\phi}^2\\
    P_2=&\frac{3\beta_1\phi^2\dot{\phi}^2}{2} + \frac{\beta_2}{2}(3\Theta\dot{\phi}^2 + \ddot{\phi}^2 +12H\dot{\phi}\ddot{\phi}+2\dot{\phi}\dddot{\phi}) \\
    P_3 =&\frac{\tilde{\xi}_1}{2}(\Theta\dot{\phi}^2 +4H\dot{\phi}\ddot{\phi}) - \frac{\tilde{\xi}_2}{2}(3\Theta\dot{\phi}^2 +2 \ddot{\phi}^2 + 12H\dot{\phi}\ddot{\phi} +2 \dot{\phi}\dddot{\phi})\\
    &+\tilde{\xi}_3( 8\dot{H}\dot{\phi}^2 + 8( 2 \dot{\phi}\ddot{H}+ \dot{H}\ddot{\phi}) \phi + 36 H^2\dot{H}\dot{\phi}^2 + 48 H\dot{H}\phi\dot{\phi} + 24 H\ddot{H}\phi^2 + (18\dot{H}^2+4\dddot{H})\phi^2)\\
    &+\tilde{\xi}_4(8H^3 \phi \dot{\phi} + 4H^2 \dot{\phi}^2 + 4H^2\phi \ddot{\phi} + 8H\dot{H}\phi \dot{\phi}]
    \end{aligned}
\end{equation}
and $(\Omega^{(6)}_{MSS})_{,a}$ can be found in Appendix \ref{appA}. It is now easy to read the pressure of the scalar field off, it is simply the right hand side of \eqref{aeom} - dropping the $(\Omega^{(6)}_{MSS})_{,a}$ terms - and is, of course, rather complicated. In the above we have split the pressure into three contributions: $P_1$ contains the usual GR pressure as well as the pressure resulting from $\phi^2R$ and $\Omega^{(4)}$; $P_2$ contains all of the higher order, minimally coupled contributions to the pressures and $P_3$ contains all of the higher order non-minimal terms.

We can now move on to find the $n$ equation of motion, which will give us the $H^2$ equation,

\begin{equation}\label{neom}
    \begin{aligned}
    3H^2 = \frac{V(\phi)}{2M^2} - \frac{\rho_1}{M^2} - \frac{\rho_2+\rho_3}{M^4} - (\Omega^{(6)}_{MSS})_{,n}
 \end{aligned}
    \end{equation}
where the densities are given by the following expressions,
\begin{equation}\label{densities}
    \begin{aligned}
        \rho_1 &= -\frac{1}{2}\dot{\phi}^2+6\tilde{\alpha}_1\dot{H}^2 + 18\tilde{\alpha}_2H^2(H^2+\dot{H}) +3\xi_1H(\phi^2H-\phi\dot{\phi}),\\
        2\rho_2 &= \beta_1\phi^3(\ddot{\phi}+\dot{\phi}H) + \beta_2(\ddot{\phi}^2 + 6\dot{\phi}\ddot{\phi}H+9\dot{\phi}^2H^2),\\
        \rho_3 &= -\frac{3}{2}\tilde{\xi}_1\dot{\phi}^2H^2 - \frac{3}{2}\tilde{\xi}_2\dot{\phi}^2\dot{H} + 6\tilde{\xi}_3\phi^2\dot{H}^2 + 6\tilde{\xi}_4\phi^2(H^4 + H^2\dot{H}),
    \end{aligned}
\end{equation}
where $(\Omega^{(6)}_{MSS})_{,n}$ can be found in Appendix \ref{appA}. We can again easily read off the density of the scalar field by taking the right hand side of the above equation and dropping the $(\Omega^{(4)}_{MSS})_{,n}+(\Omega^{(6)}_{MSS})_{,n}$ terms. Again, the form of this equation is rather complicated but it encompasses the most general non-minimally coupled field equations for an FLRW background. It is now quite easy to read off the usual GR equation of state parameter and it is clear that the above pressures and densities do indeed recover this important limit.
Next we turn to the scalar field equations of motion, this is given by,

\begin{equation}\label{compsfe}
    \ddot{\phi} + 3 H \dot{\phi} -\left(\frac{1}{2}\frac{\partial V}{\partial \phi}\right) + 6 \xi_1 X_{\xi_1} = \frac{3}{M^2}\left(\beta_1 X_{\beta_1} + \beta_2 X_{\beta_2} + \tilde{\xi}_1 \tilde{X}_2 + \tilde{\xi}_2 \tilde{X}_3 + \frac{(7 \tilde{\xi}_2 - 3 \tilde{\xi}_1}{2}\tilde{X}_4 \right) + \frac{12}{M^2}\phi\mathcal{M}_\phi,
\end{equation}

where,
\begin{equation}
    \begin{aligned}
    X_{\xi_1}&=\phi\bigg(2H^2+\dot{H}\bigg)\\
    X_{\beta_1}&= \phi(3H\phi\dot{\phi}+\dot{\phi}^2+\phi\ddot{\phi})\\
    X_{\beta_2}&= (3H\dot{H}\dot{\phi}+\dot{\phi}\ddot{H}+3H^2\ddot{\phi}+2\dot{H}\ddot{\phi}+2H\dddot{\phi}+\tfrac{1}{3}\phi^{[4]})\\
    \tilde{X}_2&=(3H^3\dot{\phi}+H^2\ddot{\phi}) \\
    \tilde{X}_3&=(\dot{\phi}\ddot{H}+\ddot{\phi}\dot{H}) \\
    \tilde{X}_4&=H\dot{H}\dot{\phi} \\
    \mathcal{M}_\phi&=\tilde{\xi}_3\dot{H}^2 + \tilde{\xi}_4\left(H^2(H^2+\dot{H})\right),
  \end{aligned}
\end{equation}
and $\phi^{[4]}$ is the fourth time derivative of the scalar field.
Once again we recover the result in \cite{scalartensor1}.

We also note that it is trivial, if somewhat tedious, to check that $\rho_\phi$ and $P_\phi$ satisfy the conservation equation for the scalar field,
\begin{equation}
    \dot{\rho}_\phi+3H(P_\phi+\rho_\phi)=0,
\end{equation}
and reproduces \eqref{compsfe}. This is an important cross check on the model presented here and confirms the accuracy of the previous equations.

\subsection{Recovering Models}
\noindent In this subsection we explore some of the known models that can be recovered from the general action used throughout this paper. If we set every coupling apart from $\xi_1$ to zero then we recover the result in \cite{scalartensor1}. In fact it is very easy to recover the pressure of one of the many models of modified gravity from this equation. For example, if we were to only consider higher order, minimally coupled scalar field terms the departure of the pressure of the scalar field from GR would become,
\begin{equation}\label{examplepressure1}
    \delta P_\phi=-\frac{1}{M^4}\big(3\beta_1\phi^2\dot{\phi}^2+\beta_2(9H^2\dot{\phi}^2+12H\dot{\phi}\ddot{\phi}+6\dot{H}\dot{\phi}^2+\ddot{\phi}^2+2\dot{\phi}\dddot{\phi})\big),
\end{equation}
and for derivative-only non-minimal coupling \`a la \cite{derivcoupling}, we set $\beta_1,\beta_2,\xi_1,\xi_4,\xi_5=0$ and obtain,
\begin{equation}\label{examplepressure2}
    \delta P_\phi=-\frac{1}{M^4}\bigg(\tilde{\xi}_3(3H^2\dot{\phi}^2+2\dot{H}\dot{\phi}^2+4H\dot{\phi}\ddot{\phi})+\tilde{\xi}_4(\ddot{\phi}^2+\dot{\phi}\dddot{\phi})\bigg).
\end{equation}
Note that in the slow-roll approximation all of these higher order corrections disappear and so one would need to modify such approximations to take account of this fact, for example $\ddot{\phi}\ll\dot{\phi},\phi$. We can also write down the $\delta\rho_\phi$ for the two examples given in \eqref{examplepressure1} and \eqref{examplepressure2}. In the first example, the change in density is easy to read off as it is simply $\rho_2$ in \eqref{densities},
\begin{equation}
    2M^4\delta\rho_\phi = \beta_1\phi^3(\ddot{\phi}+\dot{\phi}H) + \beta_2(\ddot{\phi}^2 + 6\dot{\phi}\ddot{\phi}H+9\dot{\phi}^2H^2)
\end{equation}
and in the second example we have,
\begin{equation}
    M^4\delta\rho_\phi = -\frac{3}{2}(\tilde{\xi}_1\dot{\phi}^2H^2 +\tilde{\xi}_2\dot{\phi}^2\dot{H}).
\end{equation} 
We can write down the scalar field equations that govern the cases that lead to \eqref{examplepressure1} and \eqref{examplepressure2}, the first of which gives,
\begin{equation}
    \ddot{\phi}+3H\dot{\phi}+\frac{1}{2}\frac{\partial V}{\partial \phi}=\frac{3}{M^2}\bigg(\beta_1X_{\beta_1}+\beta_2X_{\beta_2}\bigg),
\end{equation}
and the latter case yields,
\begin{equation}
    \ddot{\phi}+3H\dot{\phi}+\frac{1}{2}\frac{\partial V}{\partial \phi}=\frac{12\phi}{M^2}\mathcal{M}_\phi.
\end{equation}
Now in the former case it is difficult, without making approximations, to determine the effective mass of the scalar field, however in the case resulting from \eqref{examplepressure2} it is very simple to read of the scalar field mass,
\begin{equation}
    m_\phi^2=m_0^2+\frac{12}{M^2}\mathcal{M}_\phi,
\end{equation}
where $m_0^2$ is the bare scalar field mass coming from the potential $V(\phi)$. This is in fact the unambiguous scalar field effective mass for the whole theory as this is the only expression that can source terms that do not contain derivatives of the scalar field. This allows us to impose some restrictions on the parameters of the theory if we make some justified approximations - if we assume the scalar field has bare mass zero, $\dot{H}\ll H$, we have a flat universe and we do not allow for effective tachyons\footnote{The lack of an effective tachyon is not \textit{a priori} required to perform cosmological calculations, and in some cases it can be useful \cite{tachyon1,tachyon2,tachyon3} but it is a nice feature to have in a theory and so these restrictions are not strictly necessary but it gives some na\"ive idea of restrictions that can be imposed based on physical and observation based arguments.} we find,
\begin{equation}
    \tilde{\xi}_4\geq0,
\end{equation}
and so based on some heuristic assumptions we are easily able to constrain the effective field theory.

\section{Inflation in the Non-Minimal Model}\label{Inflation}
\noindent
Using the previously found pressure and density we are able to find the equation of state parameter $w_\phi$ of the scalar field,
\begin{equation}
    w_\phi=\frac{P_\phi}{\rho_\phi},
\end{equation}
which is well known, although quite complicated in the full non-perturbative case, the same complexity appears in the $M^{-2}$ expansion. The order $M^0$ piece of the general equation of state parameter is given by terms involving $\xi_1$ and GR terms only and is given by,
\begin{equation}
    w_\phi^0 = \frac{\dot{\phi}^2 - V + 2\xi_1 \left( \Theta \phi^2 + 2 \dot{\phi}^2 + (4 H \dot{\phi} + 2 \ddot{\phi})\phi   \right)}{\dot{\phi}^2 + V  +6\xi_1H(\phi^2H-\phi\dot{\phi})}
\end{equation}
Of course this is not particularly instructive, and in fact the general equation of state parameter is rather useless owing to its unwieldy nature. However, later in this section we shall use a simple approximation which we call, rather grandly, \textit{The Age of the Universe Approximation}. This amounts to assuming an effectively constant Hubble parameter $H$, such that all derivatives of it disappear, and assuming that the variation of the scalar field takes place on orders of the age of the universe. It is a rather coarse set of assumptions but given that we are taking a late-time effective field theory approach to cosmology here, it is a set of assumptions that will hold with some validity in the current universe.

We now briefly assess the two models used as examples above for completeness. Looking at the minimally coupled model, first presented in \eqref{examplepressure1}, we take the large $M$ limit of the equation of state as this draws out much of the effective physics. Up to order $M^{-2}$ in the expansion the equation of state parameter is given by,
\begin{equation}\label{wforbeta}
    \begin{aligned}
    w_\phi\approx& w_\phi^{GR}+\frac{1}{M^2(V+\dot{\phi}^2)^2} [\left(\dot{\phi}^2-V\right) \left(\ddot{\phi} \left(\beta _1 \phi ^3+\beta _2 \ddot{\phi}\right)+9 \beta _2 H^2 \dot{\phi}^2+H \dot{\phi} \left(\beta _1 \phi ^3+6 \beta _2 \ddot{\phi}\right)\right)\\
    &\left(\dot{\phi}^2-V\right) \left(\beta _2 \left(\left(9 H^2+6 \dot{H}\right) \dot{\phi}^2+12 H \dot{\phi} \ddot{\phi}+\ddot{\phi}^2+2 \dddot{\phi} \dot{\phi}\right)-3 \beta _1 \phi ^2 \dot{\phi}^2\right)],
    \end{aligned}
\end{equation}
where $w_\phi^{GR}=(\dot{\phi}^2-V)/(\dot{\phi}^2+V)$ is the usual minimally coupled GR result. As for the second example in \eqref{examplepressure2}, the order $M^{-2}$ equation of state parameter is given by,
\begin{equation}
    \begin{aligned}
        w_\phi &\approx w^{GR}_\phi  + \frac{(\dot{\phi}^2-V)}{M^2(V+\dot{\phi}^2)^2}\big(3\dot{\phi}(\tilde{\xi}_1\dot{\phi}H^2 + \tilde{\xi}_2\dot{H})\big)\\
        &-\frac{1}{M^2(V+\dot{\phi}^2)}\Big(\tilde{\xi}_3\dot{\phi}(3\dot{\phi}H^2 + 2\dot{\phi}\dot{H} + 4\ddot{\phi}H) + \tilde{\xi}_4(\ddot{\phi}^3+\dot{\phi}\dddot{\phi}))\Big).
    \end{aligned}
\end{equation}
We can turn the above results into a constraint on some of the model parameters. Recall that for a dominant fluid species present in the universe to cause acceleration, we require that
\begin{equation*}
    \omega_{\phi} < -\frac{1}{3}.
\end{equation*}
This relation for the $\xi_1$ only terms yields the following condition on $\xi_1$ for the fluid to cause an accelerated expansion,
\begin{equation}\label{accelcond}
    \xi_1 < \frac{2 \dot{\phi}^2 - V}{3(\Theta \phi^2 + 2 \dot{\phi}^2 + (4 H \dot{\phi} + 2 \ddot{\phi}) \phi + H (\phi^2 H - \phi \dot{\phi}))}.
\end{equation}
By choosing some specific potential one can obtain a numeric constraint on $\xi_1$.

\subsection{Slow-Roll Approximation}
\noindent
In this subsection we investigate the slow-roll approximation applied to our model. Slow-roll approximations are well known in the standard cosmological context and have been investigated for a number of models, see \cite{slowroll1,slowroll2,slowroll3} and references therein. The non-minimal slow-roll parameters in the Jordan frame have been analysed in \cite{slowroll2,slowroll3}, however we shall have to be slightly more cavalier with the restrictions in order to work with the more complex model presented in this paper. The first thing we shall do is to analyse the parameters,
\begin{equation}\label{slowrollparams}
    \epsilon = -\frac{\dot{H}}{H^2}, \qquad \tilde{\epsilon} = \frac{\dot{\epsilon}}{H\epsilon},
\end{equation}
which gives the usual slow-slow roll parameters in GR. 
The first slow roll parameter up to $M^{0}$ in the $M$ expansion is

\begin{equation}
    \epsilon = 3 \frac{(2 \dot{H} \xi_1 \phi^2 + (2 \xi_1 -1)\dot{\phi}^2 + 2 \xi_1 \phi(\ddot{\phi} - H \dot{\phi}) + 4 \tilde{\alpha}_1(\dddot{H} + 3 H \ddot{H} + 6 \dot{H}^2)   } {V + 12 \tilde{\alpha}_1 \dot{H}^2 -6 H^2 (\xi_1 \phi^2 + 12 \tilde{\alpha}_1 \dot{H}) - \dot{\phi}^2 -12H (\xi_1 \phi \dot{\phi } + 2 \tilde{\alpha}_1 \ddot{H})}.
\end{equation}
The subsequent terms are too complex to show here unless one makes some specific model choices. The next to leading order term that appears is proportional to $M^{-2}$ which can be found in the Mathematica notebook accompanying this paper. One can notice that in this expression only the non-minimal coupling term $\xi_1$ and the dimension-4 curvature operators appear at this order in the $M$ expansion. Taking the limit in which all coupling constants go to zero we obtain 
\begin{equation*}
    \epsilon^{GR} =3\left( \frac{ \dot{\phi}^2}{\dot{\phi}^2 - V}\right),
\end{equation*}
which is the usual GR result. The second slow roll parameter is 

\begin{flalign}
        \tilde{\epsilon} = 6 \left(\frac{2 \xi_1 \dot{H} \phi^2 + (2 \xi_1 -1)\dot{\phi}^2 + 2 \xi_1 \phi \ddot{\phi} + 4 \tilde{\alpha}_1 (6 \dot{H}^2 + \dddot{H})} {V - \dot{\phi}(12 \xi_1 H \phi + \dot{\phi}) + 12 \tilde{\alpha}_1 (\dot{H}^2 - 2 H \ddot{H})  } \right) +&& \\ \nonumber
        2 \left(\frac{\xi_1 \ddot{H}\phi^2 + (3 \xi_1 -1)\dot{\phi} \ddot{\phi} + H (6 \tilde{\alpha}_1 \dddot{H} -\xi_1 \dot{\phi}^2) + \xi_1 \phi(\dot{H}\dot{\phi} - H \ddot{\phi} + \dddot{\phi}) + 2 \tilde{\alpha}_1 (15 \dot{H}\ddot{H} + H^{(4)})}{H(2 \xi_1 \dot{H}\phi^2 + (2 \xi_1 -1) \dot{\phi}^2 + 2 \xi_1 \phi(\ddot{\phi}- H \dot{\phi}) + 4 \tilde{\alpha}_1 (6 \dot{H}^2 + 3 H \ddot{H} + \dddot{H})} \right) &&.
\end{flalign}
Again, this is quite an unwieldy expression to work with, but if we take the GR limit where all coupling constants go to zero we obtain
\begin{equation}
    \tilde{\epsilon}^{GR} = 2 \left(\frac{\ddot{\phi}}{H \dot{\phi}} \right)+6\left( \frac{ \dot{\phi}^2
}{\dot{\phi}^2 - V}\right).
\end{equation}
We conclude this section by mentioning that the slow roll parameters can be connected to the observables $(n_s, r)$ via the following relations \cite{Forconi:2021que}:
\begin{equation}
    n_s -1 = -2 \epsilon- \tilde{\epsilon},
\end{equation}
and 
\begin{equation}
    r = 16 \epsilon.
\end{equation}
Here $r$ is the tensor to scalar ratio and $n_s$ is the scalar spectral index. By specifying some specific choice of model parameters one can compare the predictions of their chosen model to that of the constraints outlined in the Planck data \cite{Planck:2018vyg}.

\subsection{Age of the Universe Approximation}
\noindent
In this subsection we take an \textit{age of the universe} approximation. In this approximation we send all derivatives of the Hubble parameter to zero and conjecture that the scale field varies over times that are of the same order as the age of the universe, this is a valid approximation in the late stage expansion of the universe. Thus in this approximation we take $\dot{\phi}\sim H\phi$ and $\dot{H},\ddot{H},...=0$ as well as,
\begin{equation}\label{modslow}
    \phi^{[n]}\sim H^n\phi.
\end{equation}
The pressures in \eqref{pressures} reduce to a very simple form in this approximation and they are given by,
\begin{equation}
    P_1 = \frac{1}{2}H^2\phi^2(22\xi_1+1), \quad P_2 = \frac{3}{2}H^2\phi^2(\beta_1\phi^2+8\beta_2H^2), \quad P_3 = \frac{H^4\phi^2}{2}(7\tilde{\xi}_1 - 25\tilde{\xi}_2 + 32\tilde{\xi}_4),
\end{equation}
and the densities in \eqref{densities} become,
\begin{equation}
    \rho_1 = \frac{H^2}{2}(-\phi^2+36\tilde{\alpha}_2H^2), \quad \rho_2 =H^4\phi^2(\beta_1 + 8\beta_2), \quad \rho_3 = \frac{H^4\phi^2}{2}(12\tilde{\xi}_4 - \tilde{\xi}_1).
\end{equation}
Of course these pressures and densities are far simpler than the full set of equations and they further allow us to explore some of the late stage behaviour of the effective model. We can take these pressures and densities and find the approximate equation of state parameter for the scalar field expanded to $M^{-2}$,
\begin{equation}\label{equationofstate}
    \begin{aligned}
    w_\phi\approx& w^{GR}_\phi +\frac{H^2\phi^2(-3\beta_1\phi^2V + q_1H^6 - 3\beta_1\phi^4H^2 + q_2VH^2 - q_3\phi^2H^4)}{M^2(V+\phi^2H^2-26\alpha_2H^4)^2}
    \end{aligned}
\end{equation}
where,
\begin{equation}
    w_\phi^{GR}=\frac{(22\xi_1-1)H^2\phi^2 - V}{V+H^2\phi^2 - 36\tilde{\alpha}_1H^4},
\end{equation}
is a slightly modified GR equation of state parameter (we make this modification since in the age of the universe approximation $(\partial\phi)^2$ contributes the same term as $\phi^2R$, that is $\phi^2H^2$). Note that $q_1,q_2,q_3$ are various combinations of the coupling constants and are not particularly important. The approximation made in \eqref{modslow} puts all time dependence of the field into the Hubble parameter and hence any variation of the fields is expected to occur on the age of the universe. 

We can the take the first example model \eqref{wforbeta} and look at a massive non-self interacting scalar field, such that $V=m^2\phi^2$, which varies very slowly such that $\dot{\phi}\sim H\phi$ and all other derivatives of $H$ and $\phi$ vanish,
\begin{equation}
    w_\phi\approx \tilde{w}_\phi^{GR}+\frac{2(\beta_1-4\beta_2)H^6-3\beta_1m^2\phi^2H^2 - H^4(2(\beta_1+20\beta_2)m^2+3\beta_1\phi^2}{M^2(H^2+m^2)^2} ,
\end{equation}
where $\tilde{w}_\phi^{GR}=(H^2-m^2)/(H^2+m^2)$. Interestingly, if we demand that the scalar field's mass is $m^2\sim H^2$, then $\tilde{w}^{GR}_\phi=0$ and the scalar field acts (approximately) as a pressure matter contribution to the action rather than a dark energy field. This approximation is the statement that the scalar field should not decay into gravitational perturbations, via the non-minimally coupled terms, on time scales less than the age of the universe. Of course, the field will behave in some way like a dark energy field but these effects are suppressed by the cut-off mass squared in this instance. Hence if we wish to have a dark energy field in any model where $\xi_1=0$, the scalar field would need to have self interactions (or interactions with other matter fields not analysed here). The above equation shows that the equation of state varies quadratically with the value of the scalar field. 
We can also find the equation of state parameter for the second example, in which we only consider derivative terms that are non-minimally coupled, in the limit outlined above the equation of state is
\begin{equation}\label{nonminislow}
    w_\phi\approx \tilde{w}_\phi^{GR}+\frac{H^4(p_1m^2 + p_2H^2)}{(m^2+H^2)^2M^2},
\end{equation}
which has no $\phi$ dependence and where $p_1 = 6\tilde{\xi}_1-25\tilde{\xi}_2+44\tilde{\xi}_4$ and $p_2 = 8\tilde{\xi}_1-25\tilde{\xi}_2+20
\tilde{\xi}_4 $. This is interesting as it shows, in this slowly varying model, that the equation of state parameter is essentially a constant no matter what value $\phi$ takes, and in the case of $V= m^2\phi^2$ very close to zero.

In this age of the universe approximation, the system becomes much simpler to solve as the differential equations reduce to polynomials in the parameters. We shall set $k=0$ as before, and take the approximation in \eqref{modslow}. The field equations \eqref{aeom}, \eqref{neom} and \eqref{compsfe} then take the form,
\begin{equation}\label{aoueqs}
    \begin{aligned}
    3H^2&=\frac{H^2}{M^2}\frac{\phi^2}{2}(1-22\xi_1)-\frac{H^4}{M^4}\bigg(\frac{3\phi^2}{2H^2}(\beta_1\phi^2+8\beta_2H^2) +\frac{\phi^2}{2}(7\tilde{\xi}_1 - 25\tilde{\xi}_2 + 32\tilde{\xi}_4) \bigg)-\frac{V}{2M^2},\\
    3H^2&=\frac{H^2}{2M^2}(\phi^2-36\tilde{\alpha}_2H^2)+\frac{H^4}{2M^4}\bigg(\phi^2(2\beta_1 + 16\beta_2+ 12\tilde{\xi}_4 - \tilde{\xi}_1) \bigg)+26\tilde{\omega}_1\frac{H^6}{M^4}+\frac{V}{2M^2}\\
    \frac{\partial_\phi V}{2\phi}&=4H^2(1+3\xi_1)-15\frac{H^2}{M^2}\phi^2 + \frac{H^4}{M^2}(16\beta_2+12\tilde{\xi}_1+12\tilde{\xi}_4)
    \end{aligned}
\end{equation}
respectively. Now, we have three unknown parameters in this model, the background value of the scalar field $\phi_0$, the Hubble parameter $H^2$ and the cut-off mass $M^2$. It is now possible to solve for some of these using the above equations for a given polynomial potential.

As an example we shall solve the scalar field equation of motion in this limit as this will give us constraints on $\xi_1$. Let us consider setting $V=-\frac{m^2}{2}\phi^2+\frac{\lambda}{4}\phi^4$ and use the last equation in \eqref{aoueqs} to find $\phi_0$ in terms of the couplings, $H$ and $M$,
\begin{equation}\label{phi0}
    \begin{aligned}
    \phi_0^2&=\frac{m^2}{\lambda}+\frac{8H^2(1+3\xi_1)}{\lambda}-30\frac{H^2}{M^2}\frac{m^2}{\lambda^2} - 8\frac{H^2}{M^2}\frac{H^2}{\lambda^2}(30+90\xi_1-\lambda(4\beta_2+3\tilde{\xi}_1+3\tilde{\xi}_4))
    \end{aligned}
\end{equation}
where we can see the usual vev of a Higgs-like scalar field as well as some gravitational corrections at $\mathcal{O}(1)$. We can use this to impose some restrictions on the parameters but demanding that $\phi_0\in\mathbb{R}$. If we work to $\mathcal{O}(1)$ we find the following condition
\begin{equation}\label{mconst}
   m\geq\sqrt{-8(1+3\xi_1)}H,
\end{equation}
and hence if we want a $m\in\mathbb{R}$ implies,
\begin{equation}
    \xi_1\leq -\frac{1}{3}
\end{equation}
which gives an approximate constraint on the value of the coupling. The constraint in \eqref{mconst} also tells us that the lower bound for the mass $m$ is proportional to the Hubble value. This should come as no surprise as this inequality basically states that the scalar field must decay on time scales around the length of the universe - an assumption implicit in the age of the universe model. We can work to higher order allowing us to impose further reality constraints on the mass of the scalar fields using this vev. At $\mathcal{O}(M^{-2})$ one finds the following correction to the scalar field mass,
\begin{equation}
    \delta_1m= \frac{H^3}{M^2}\sqrt{\frac{-2}{(1+3\xi_1)}}(4\beta_2+3\tilde{\xi}_1+3\tilde{\xi}_4),
\end{equation}
which implies the two following facts. In this limit $\xi_1\neq-\frac{1}{3}$ and so the above inequality becomes strict in the sense that $\xi_1<-1/3$. This further implies that it is not possible for the mass of the scalar field to be 0 unless $H=0$ since setting the order 1 part of $m=0$ will introduce a pole in the order $M^{-2}$ part. We can find further constraints on $\xi_1$ by considering cosmological parameters in this regime. In this approximation, the first order part of the cosmological observable $n_s$ becomes,
\begin{equation}
    n_s-1 = -2-6H^2\phi^2\bigg(\frac{(4\xi_1-1)}{V-(12\xi_1+1)H^2\phi^2}-\frac{(2\xi_1-1)}{V-(18\xi_1+1)H^2\phi^2}\bigg),
\end{equation}
which according to cosmological data \cite{Planck:2018vyg} implies $n_s-1\sim0$ and using this condition the above equation can be solved for the $\xi_1$ coupling to give,
\begin{equation}
    \xi_1\sim\frac{17 H^2 \phi_0 ^2-2 V\pm\sqrt{3} \sqrt{115 H^4 \phi_0 ^4-44 H^2 V \phi_0 ^2+4V^2}}{24 H^2 \phi_0 ^2}.
\end{equation}
Now, taking the Higgs-like potential in the above example, as well as the vev solution in \eqref{phi0}, one finds the following solution,
\begin{equation}
    \xi_1\sim \frac{H^2 f_1(\lambda)+ f_2(\lambda) m^2+\lambda^2\sqrt{3} \sqrt{ 4 H^4 f_3(\lambda)+4 H^2 f_4(\lambda) m^2+f_5(\lambda)}}{24 H^2 \left(2 \lambda ^4+2 \lambda ^2-1\right)}
\end{equation}
where $f_i(\lambda)$ are various polynomials in the coupling parameter $\lambda$ - these are not important for the analysis here. However, note that in this regime one can, at least at this order, impose some constraints on the value of $\lambda$ owing to the potential pole arising from the denominator of the above. Finally, one can consider the constraint in \eqref{accelcond} which imposes accelerated expansion. For the potential outlined above, the constraint becomes,
\begin{equation}
    \xi_1<\frac{2}{33} + \frac{1}{66}\frac{m^2}{H^2}-\frac{\lambda}{132}\frac{\phi^2}{H^2}.
\end{equation}
Now we can insert the perturbative solution for $\phi_0$ in \eqref{phi0} to find,
\begin{equation}
    \xi_1 < \frac{m^2}{156H^2},
\end{equation}
but we assume that $m^2>0$ and $H^2>0$, hence this bound is trivially satisfied by the reality of $m$ from above. Hence, in the age of the universe approximation, with a Higgs like potential, the cosmology is an accelerating one.

\subsection{$\mathbb{Z}_2$ Symmetry Breaking}
\noindent
In this subsection we briefly discuss the consequences of breaking the $\mathbb{Z}_2$ symmetry of the scalar field to a constant vacuum expectation value $\phi_0$. This symmetry breaking will induce a cosmological constant in the action \eqref{CA} whose value will be determined by solving $\partial_\phi V=0$ and the value of the constant $\Lambda$ will be given by,
\begin{equation}
    \Lambda=V(\phi_0).
\end{equation}
If one assumes that the $\phi_0$ is truly constant ($\dot{\phi_0}=0$) then one has to look at \eqref{CSFE} which reduces to,
\begin{equation}\label{SSB}
    M^2R+\frac{\xi_4}{\xi_1}R_{\mu\nu}R^{\mu\nu}+\frac{\xi_5}{\xi_1}R_{\mu\nu\rho\sigma}R^{\mu\nu\rho\sigma}+\frac{\xi_6}{\xi_1}=0,
\end{equation}
which places non-trivial constraints on theory. If one substitutes \eqref{FLRW} into the above equation one finds that the theory must obey,
\begin{equation}\label{z2sym}
    \frac{M^2\xi_1}{2}\bigg(\dot{H}+2H^2\bigg)=2\tilde{\xi}_4(H^4+\dot{H}^2)+\tilde{\xi}_3H^2\dot{H}.
\end{equation}
This equation can be satisfied if $\xi_1\sim H^2 M^{-2}$ which is not surprising given the form of \eqref{SSB} which shows that the $\xi_1$ term is $M_{Pl}^{2}$ larger than the other terms.
In line with the analysis above, whereby we took $\dot{H}\sim0$, one finds that the equation that needs to be satisfied is,
\begin{equation}
    M^2\xi_1\approx2\tilde{\xi}_4H^2,
\end{equation}
which can allows us to relate terms within the EFT via,
\begin{equation}
    \tilde{\xi}_4\approx\frac{M^2\xi_1}{2H^2}.
\end{equation}
This subsection highlights an important result concerning the generation of a cosmological constant, sourced from the scalar field, in this type of theory: we must have a very small value for $\xi_1$ in to generate a constant vacuum expectation value which in turn gives the cosmological constant. Since we have assumed that $\phi$ is constant in this derivation all of the $\beta_i$ terms along with the $\xi_2$ and $\xi_3$ terms vanish on the background solution and such terms will only be manifest once one perturbs the theory. One could of course not impose the constant nature of $\Lambda$ \cite{Gorbar:2003yp} and simply find a `solution' to $\delta S/\delta\phi=0$ and use this as the source for the cosmological constant but this would introduce further curvature tensors to the action and it is not clear whether this a physical procedure - as it is unclear how to define the scalar field potential in this case. It may also be ambiguous due to integration by parts on the higher order scalar field terms. Some heuristic calculations showed that if one proceeds along this path then once the vacuum expectation value for the scalar field (which is assumed to be dynamic) has been subbed back into the action one obtains the same constant cosmological constant and simply rescales all of the other quantities in the theory at the cost of eliminating $\phi$ in the background theory as usual. If one assumes all terms that are not the kinetic operator for the scalar field are part of this generalised potential then one has to invert an operator in order to solve for $\phi$ and this requires a low momentum assumption for the value of the scalar field.

\section{Conclusions}\label{conclusions}
\noindent
In this work we have presented the most general non-minimally coupled $\mathbb{Z}_2$ symmetric field theory up to dimension six in the operators, see \eqref{CA}. This field theory was presented in the bottom-up context, however the results presented above generalise to any such scalar-tensor theory of gravity up to dimension six in the operators. The covariant field equations for this theory were presented in \eqref{CSFE}, \eqref{CEFE} and \eqref{miniSET} in Section \ref{Model} and as is typical for this class of theories they are non-linear and coupled. In general one cannot find closed, non-perturbative, analytic solutions for such equations and so one typically has to find perturbative solutions in terms of some expansion parameter. In Section \ref{Equations} the mini superspace action was presented \eqref{MSSA} for the FLRW metric with general curvature $k$ and the equations of motion were found. These equations 
of motion gave the pressure and density of the non-minimally coupled scalar field - although we did not display these explicitly in the main body of the text. Some examples were presented for $\delta P_\phi$ and $\delta\rho_\phi$ as well as the equation of state parameters in both the slow-roll approximation and the so-called age of the universe approximation. In this approximation we were able to find a number of constraints on the $\xi_1$ which led to the conclusion that $\xi_1<-1/3$ in the Higgs potential model. In fact, this requirement imposed that the universe be expanding. This follows from the constraint on the equation of state parameter in \eqref{accelcond}. We also briefly discussed the physics of $\mathbb{Z}_2$ symmetry breaking of the scalar field. It was noted that for a stationary scalar field vev, certain non-trivial conditions would have to be satisfied by the EFT such as those in \eqref{z2sym}. We then explored a number of simple inflationary models in both the slow-roll and age of the universe approximations. The exploration carried out in this paper presents very general results which cover a large range of effective theories of cosmology as well as imposing some constraints on these theories. It would be fruitful to undertake further investigations into the speed of propagation of the scalar field and gravitation waves through the FLRW-like background in order to further constrain the coefficients in such theories.

\appendix
\section*{Acknowledgements}
\section{Dimension-4 and -6 Curvature Operators}\label{appA}
We split the $\Omega$ in \ref{CA} into the dimension-4 terms $\Omega^{(4)}$ and the dimension-6 terms $\Omega^{(6)}$ where the former have coupling constants $\alpha_i$ and the latter have $\omega_i$ as their couplings. The dimension-4 operators can be chosen to be,
\begin{equation}
    \Omega^{(4)}=\alpha_1R^2+\alpha_2R_{\mu\nu}R^{\mu\nu},
\end{equation}
there are only two independent choices at this level since the Gauss-Bonnet term is a non-dynamical topological term, hence only two of the possible combinations are independent. One can choose other combinations - including the Weyl tensor \cite{speedofgravity} - but only two such terms will ever appear at this order. The dimension-6 curvature operators are well known and can be expressed as \cite{speedofgravity,Dim6ops},
\begin{equation}\label{dim6curv}
    \begin{aligned}
    \Omega^{(6)}=&\omega_1R\Box R+\omega_2 R^{\mu\nu}\Box R_{\mu\nu}+\omega_3R^3+\omega_4RR^2_{\mu\nu}+\omega_5RR^2_{\mu\nu\rho\sigma}+\omega_6R_{\mu\nu}^3+\omega_7R^{\mu\nu}R^{\rho\sigma}R_{\mu\rho\nu\sigma}\\
    &+\omega_8R^{\mu\nu}R_{\mu\rho\sigma\gamma}\tensor{R}{_\nu^{\rho\sigma\gamma}}+\omega_9R^{\mu\nu\rho\sigma}R_{\mu\nu\alpha\beta}\tensor{R}{_{\rho\sigma}^{\alpha\beta}}+\omega_{10}\tensor{R}{^\mu_\rho^\nu_\sigma}\tensor{R}{^\rho_\alpha^\sigma_\beta}\tensor{R}{^\alpha_\mu^\beta_\nu}
    \end{aligned}
\end{equation}
and so there are ten independent terms at this order. This number of terms is the major reason why certain expressions are not displayed in this work - the equations become very unwieldy and intractable very quickly once one begins to go to higher orders in the curvature operators. We can display the MSS terms that result from these operators, in a similar fashion to \eqref{dim4SSA},

\begin{equation}\label{dim6SSA}
    \begin{aligned}
    M^4\Omega_{MSS}^{(6)}=-6\bigg[
    &\frac{1}{n^6}(2\tilde{\omega}_1 H^6+3\tilde{\omega}_2H^4\dot{H}+\tilde{\omega}_3 \dot{H}^3-21\tilde{\omega}_4 H^3\ddot{H}-2\tilde{\omega}_5 \dot{H}\dddot{H}-6(\tilde{\omega}_4 +\tilde{\omega}_5)H\dot{H}\ddot{H}
+H^2[\tilde{\omega}_6 \dot{H}^2-3\tilde{\omega}_4 \dddot{H}])\bigg]
    \end{aligned}
\end{equation}

where the $\tilde{\omega}$ are given by,
\begin{equation}
    \begin{aligned}
    \tilde{\omega}_1&=144 \omega _3+36 \omega _4+24 \omega _5+9 \omega _6+6 \omega _8+4 \omega _9+2 \omega _{10}, \\
    \tilde{\omega}_{2}&=-48 \omega _1-12 \omega _2+144 \omega _3+36 \omega _4+24 \omega _5+9 \omega _6+6 \omega _8+4 \omega _9+2 \omega _{10}, \\
    \tilde{\omega}_{3}&=-24 \omega _1-6 \omega _2+36 \omega _3+12 \omega _4+12 \omega _5+5 \omega _6+4 \omega _8+4 \omega _9, \\
    \tilde{\omega}_{4}&= 4 \omega _1+\omega _2,\\
    \tilde{\omega}_{5}&=3 \omega _1+\omega _2, \\
    \tilde{\omega}_{6}&=-120 \omega _1-26 \omega _2+216 \omega _3+60 \omega _4+48 \omega _5+18 \omega _6+14 \omega _8+12 \omega _9+3 \omega _{10}.
    \end{aligned}
\end{equation}
We can then use \eqref{dim6SSA} to find the final parts of the components equations of motion: for \eqref{aeom},
\begin{subequations}
\begin{equation}
\begin{aligned}
M^4(\Omega^{(6)}_{MSS})_{,a}=\frac{6}{M_{Pl}^2}\bigg[30\tilde{\omega}_1H^4\dot{H}+2\tilde{\omega}_5H^{[5]}-3(\tilde{\omega}_3+3\tilde{\omega}_5)(\ddot{H}^2+\dot{H}H^3)
&-(63\tilde{\omega}_4+\tilde{\omega}_6)(\dot{H}^3+4H\dot{H}\ddot{H}+H^2\dddot{H})\bigg]
\end{aligned}
\end{equation}
\end{subequations}

and for \eqref{neom},
\begin{subequations}
\begin{equation}
    \begin{aligned}
M^4(\Omega^{(6)}_{MSS})_{,n}=&\frac{6}{M_{Pl}^2}\bigg[6\tilde{\omega}_1H^6+3\tilde{\omega}_3\dot{H}^3+9H^4 \tilde{\omega}_2\dot{H}-63\tilde{\omega}_4H^3\ddot{H}
&-6H\ddot{H}\bigg(3(\tilde{\omega}_4+\tilde{\omega}_5)\dot{H}\bigg)-6\dot{H}\tilde{\omega}_5\dddot{H}\bigg)\\
&+H^2\bigg(3(\tilde{\omega}_6\dot{H}^2-3\tilde{\omega}_4\dddot{H}\bigg)\bigg].
    \end{aligned}
\end{equation}
\end{subequations}

\bibliography{bibliography}
\end{document}